\documentclass[12pt]{article}
\usepackage{amsmath,amsfonts,amssymb}
\usepackage{epsfig}
\usepackage{paralist}
\usepackage{graphicx}
\usepackage{color}

\makeatletter\@addtoreset{equation}{section}\makeatother

\newtheorem{theorem}{Theorem}[section]
\newtheorem{lemma}[theorem]{Lemma}

\newenvironment{proof}[1][Proof]{\begin{trivlist}
\item[\hskip \labelsep {\bfseries #1}]}{\end{trivlist}}

\newcommand{\qed}{\hfill \ensuremath{\Box}}
\def\L{\Lambda}
\def\tk{\tilde{k}}
\def\tbeta{\tilde{\beta}}

\def\bF{\mathbb{F}}
\def\bS{\bar{S}}

\def\l{\lambda}

\def\tbF{\tilde{\mathbb{F}}}

\def\bI {\mathbb{I}}

\def\ep{\epsilon}

\def\be{\begin{equation}}
\def\ee{\end{equation}}
\def\bea{\begin{eqnarray}}
\def\eea{\end{eqnarray}}

\newcommand{\p}{\partial}
\newcommand{\mf}{\mathcal{F}}

\newcommand{\tr}{{\rm tr\,}}

\newcommand{\nn}{\nonumber}

\newcommand{\la}{\langle}
\newcommand{\ra}{\rangle}
\newcommand{\nc}{\newcommand}
\nc{\cH}{\mathcal{H}}

\renewcommand{\title}[1]{\vbox{\center\LARGE{#1}}\vspace{5mm}}
\renewcommand{\author}[1]{\vbox{\center#1}\vspace{5mm}}
\newcommand{\address}[1]{\vbox{\center\em#1}}
\newcommand{\email}[1]{\vbox{\center\tt#1}\vspace{5mm}}

\begin{document}
\begin{titlepage}
\begin{center}
\hfill \\
\hfill \\
\vskip 1cm

\title{Equilibration of Small and Large Subsystems\\
in Field Theories and Matrix Models}

\author{Nima Lashkari}

\address{Department of Physics and Astronomy, University of British Columbia\\
6224 Agricultural Road, Vancouver, B.C., V6T 1W9, Canada}

\email{nima@phas.ubc.ca}

 \end{center}
 
\abstract{It has been recently shown that small subsystems of finite quantum systems generically equilibrate.  
We extend these results to infinite-dimensional Hilbert spaces of field theories and matrix models. 
We consider a quench setup, where initial states are chosen from a microcanonical ensemble of finite energy in free theory, and then 
evolve with an arbitrary non-perturbative Hamiltonian.
Given a dynamical assumption on the expectation value of particle number density, we prove that small subsystems reach equilibrium at the level of quantum wave-function, and 
with respect to all observables. The picture that emerges is that at higher energies, larger subsystems can reach equilibrium. For bosonic fields on a lattice, in the 
limit of large number of bosons per site, all subsystem smaller than half equilibrate. In the Hermitian matrix model, by contrast, this occurs in the limit of large energy 
 per matrix element, emphasizing the importance of the $O(N^2)$ energy scale for the fast scrambling conjecture. Applying our techniques to continuum field theories on compact spaces, 
 we show that the density matrix of small momentum-space observables equilibrate. 
 Finally, we discuss the connection with scrambling, and provide a sufficient condition for a time-independent Hamiltonian to be a scrambler in terms of the entanglement entropy of its energy eigenstates.}
\end{titlepage}

\tableofcontents

\section{Introduction}

It is generally accepted that most systems equilibrate even if they are perturbed far away from equilibrium.
This is to say that the information about the initial state spreads out such that after some time all observables 
restricted to small subsystems become almost independent of time and the initial state.  
Our current understanding of equilibration in field theories remains at the level of perturbation theory, either
at small coupling or large $N$ expansion. Perturbation theory enables us to keep track of the time dependence of
  the expectation value of observables with a small number of field operators, what we refer to as small observables. However, the formalism 
  quickly becomes cumbersome as we look at large observables with many field insertions. Knowledge of all  
   observables is required to specify the quantum state of field theory. It is not clear whether the 
   effective state one infers from the expectation value of small observables is a good description of the system. 
   
   In this work, we address equilibration at the level of subsystems' density matrices 
   non-perturbatively. We argue, quite generally, that the reduced density matrix of small subsystems becomes 
   indistinguishable from the equilibrium state with respect to all observables living in the subsystem Hilbert space.
   The techniques we employ in this work apply to Hamiltonians with discrete spectrum as in
    lattice field theory, continuum field theories on compact manifolds and matrix models. 
   Our results suggest that at higher energies larger subsystems can equilibrate and in the limit of infinite energy, 
  on a lattice and in matrix models, any subsystem smaller than half the size of the system equilibrates. 
   
Our work is a generalization of quantum information techniques \cite{citeulike:3788627,reimann2010canonical,citeulike:9264197} that have been recently used to 
discuss equilibration of finite systems to the case of infinite dimensional Hilbert spaces.\footnote{For a discussion of equilibration in integrable systems see \cite{barthel2008dephasing}.} Motivated by 
 the fast scrambling conjecture \cite{Sekino:2008he}, this generalization enables us to address the question of the equilibration of
  large subsystems in field theories and matrix models. We consider a quench setting where the initial state
   is chosen from a free theory, and evolved with an arbitrary non-perturbative interacting Hamiltonian. With some reasonable physical assumptions 
    we prove that small subsystems always equilibrate. Furthermore, on a lattice of $N$ sites, initial states of energy larger than $O(N)$ equilibrate on all subsystems 
    larger than half. In the Hermitian matrix model the large-scale equilibration occurs for energies larger than $O(N^2)$. 
In continuum field theories, our results imply that there are scales $\mu_1$ and $\mu_2$, where the infra-red  and ultra-violet density matrices constructed by 
tracing out all momentum modes, respectively, higher than $\mu_1$ and lower than $\mu_2$, are almost non-dynamical and time-independent. In a quench setting from 
an eigenspace of large energy $\Delta$ in a $d$-dimensional free theory, we find that both $\mu_1$ and $\mu_2$ are $O(\Delta^{1/d})$.

In spite of the technical details of proofs, the equilibration results are based on a rather simple 
  picture. The initial state of the system lives in a large Hilbert space and contains a significant 
  amount of information. 
  If the Hilbert space that the subsystem explores is small, then it can not contain all the information about the 
  initial state and quickly loses track of it. An interesting question is what happens if one tries to 
   encode a message in the initial state. We say that the message is scrambled if it cannot be 
   retrieved from the state of any subsystem smaller than half the size of the system \cite{Sekino:2008he,Lashkari:2011yi}. With this definition, scrambling is a strong form of equilibration in which the state of large subsystems  
   become independent of initial states including atypical ones such as a tensor product states. 
    Hamiltonians, which are very powerful in generating entanglement, leave almost no room 
    in the Hilbert space of subsystems, even the large ones. 
     We will discuss the implication of this for scrambling of time-independent Hamiltonians.

    In section 2, we start by defining what we mean by equilibration. Our method closely follows the arguments and approach to equilibration developped in the 
    quantum information community over the past few years. Therefore, we devote section 3 to a quick review of results in finite systems.
    Our main results appear in section 4 which generalizes equilibration theorems to infinite dimensional Hilbert spaces in quench 
    settings and includes as an example the equilibration in the following three systems: bosons on a lattice,  the Hermitian matrix model and
     field theory in momentum space. All results up to section 5 hold for generic interactions. The dynamical input from the Hamiltonian appears in 
     section 5 where we expand on the connection with scrambling.


%


\section{Basic setup}

    The many-particle quantum systems we are interested in are described by a set of physical states
  living in a tensor product of Fock spaces: $\psi\in\mf= \mf_1\otimes\mf_2\otimes\hdots\mf_N $. The 
  tower of states in Fock spaces $\mf_i$ are generated by the creation operators $a_i^\dagger$. In particular we 
  consider three systems described in this manner: 
 \begin{enumerate} 
  \item Bosons on a lattice where $a^\dagger_i$ creates a boson at site $i$.
    
 \item An $N\times N$ Hermitian matrix model which is equivalent $N$ interacting fermions. The creation operator $a_i^\dagger$ 
  increases the energy of $i^{th}$ fermion by one unit. 
%

  \item Quantum fields in momentum space on a compact space where $a_k^\dagger$ creates
  a particle with momentum $k$. 
  \end{enumerate}

  Note that for bosons on a lattice and quantum fields in momentum space, any $\psi\in \mf$ is physical, whereas in the matrix model the physical 
  states are only the ones invariant under gauge transformations. In the Hermitian matrix model we consider here, due to
  the fermionic nature of degrees of freedom, the physical states are anti-symmetrized over 
  $N$ sites \cite{brezin1978planar}.

      In physical problems of interest, there are typically global constraints on the Hilbert space of initial
states. They are chosen from a subspace $\cH_R$ of $\mf$. 
Note that the restriction is imposed only on initial states and the wave-function can leave $\cH_R$ as it evolves. 
We denote the subspace the time-evolved initial states explore by $\cH_T$. It is the subspace spanned by energy
eigenstates with nonzero projection into $\cH_R$. 

Consider an initial physical state  $\psi\in \cH_R\subset \mf$. 
As $\psi$ evolves in time, if the system equilibrates with respect to subsystem $S$, the reduced density matrix on $S$ becomes close to an equilibrium state $\rho_S^{equil}$. 
    Any comparison of distances between physical states requires a choice of metric 
  in the Hilbert space. Trace norm provides a 
  metric with a natural operational meaning. If two states $\rho_S$ and $\rho_S^{equil}$ are close in trace norm, i.e. $\|\rho_S-\rho_S^{equil}\|_1<\epsilon$, then any
  measurement to distinguish them succeeds with probability at most $\frac{1}{2}+\epsilon$.\footnote{For a basic reference on trace norm, and other commonly used measures of distance in Hilbert space see \cite{nielsen2010quantum}.}
For some choice of Hamiltonians and initial states it is possible that  
  that $\rho_S(t)$ approaches its equilibrium state monotically in time \cite{PhysRevLett.100.030602,PhysRevLett.106.050405}.
    We say $\rho_S$ equilibrates in a {\it strong} sense, if for any $\ep>0$ there exists a $\tau$ such that for all $t>\tau$, $\|\rho_S(t)-\rho_S^{equil.}\|_1<\epsilon$. 
    More generically, systems tend to equilibrate in a {\it weak} sense: $\rho_S(t)$ becomes close to its equilibrium 
    value, and spends most of its time in a small neighbourhood around it. We say $\rho_S$ equilibrates 
     in a {\it weak} sense, if there exists an $\epsilon\ll 1$ such that:
     \bea\label{norm}
     \langle \|\rho_S(t)-\rho_S^{equil}\|_1\rangle_t\leq \ep,
     \eea
     where $\la \sigma(t)\ra_t=\lim_{T\to\infty}\frac{1}{T}\int_0^T\sigma(t)dt$; see figure 1.
 \begin{figure}\label{fig1}
\begin{center}
\input{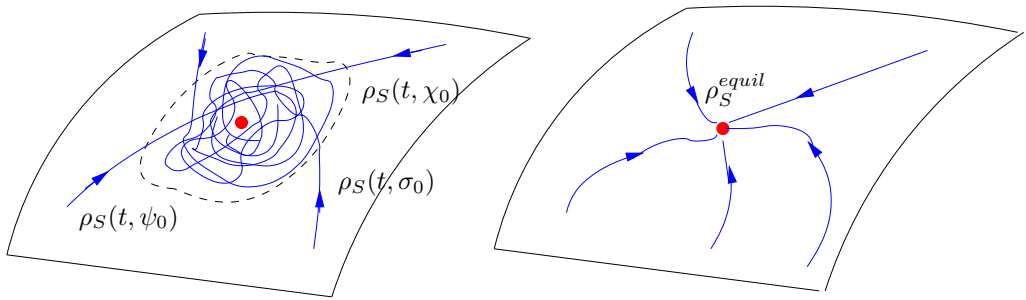}
\caption{This figure schematically shows the difference between weak and strong equilibrations. 
As the subsystem density matrix evolves it comes close to its equilibrium value with respect to the trace norm.
 In weak equilibration (left-hand side) $\rho_S(t)$ spends most of its time in an $\ep$ neighborhood of the 
  equilibrium state which is independent of the initial state $\rho_0$, $\phi_0$ or $\chi_0$. This is in contrast with 
  strong equilibration (right-hand side) where equilibrium is approached monotically in time.}
  \end{center}
\end{figure}
     
     Conceptually, the weak equilibration matches our 
     intuition of equilibration better, since it leaves the possibility for $\rho_S$ to fluctuate 
     away from equilibrium in a time-independent way. The bound in (\ref{norm}) is the statement that large fluctuations are rare. 
          
     We would like to stay as generic as possible; therefore we focus on weak equilibration.
Following \cite{citeulike:3788627}, our strategy is to show first that the distance between a subsystem's density matrix and
 its time-averages $\omega\equiv\lim_{T\to\infty}\la\rho_S(t)\ra_t$ is bounded by a constant:
     \bea
     \langle \|\rho_S(t)-\omega_S\|_1\rangle_t\leq \eta_S.
     \eea
 Then, we show that for most initial states in the initial ensemble the constant $\eta_S$ is small. This
  establishes the time-independence of the subsystem state. 
 In principle $\omega_S$ is a function of the initial state $\psi$. Next, 
we need to show that $\omega_S(\psi)$ is well-approximated by its average over all $\psi\in\cH_R$:
\bea
 \la\|\omega_S(\psi)-\la\omega_S\ra_\psi\|_1\rangle_\psi\leq \ep'.
\eea

This is the statement that the equilibrium state keeps almost no memory of the initial perturbation.
However, this does not necessarily imply that the equilibrium state is well approximated by the Gibbs state. In cases where 
    the asymptotic state is the Gibbs state we say that the system has thermalized. It is known that thermalization is
    not as ubiquitous in nature as equilibration. There are many known examples of systems and initial states that 
    equilibrate but never thermalize \cite{PhysRevLett.106.050405,rigol2009breakdown}. It is a trivial extension of the above definitions to introduce notions of
     weak or strong scrambling or thermalization.
 
Another noteworthy comment is the difference between the equilibration of subsystems as defined above and
 equilibration with respect to coarse-grained observables \cite{reimann2010canonical, citeulike:9264197, PhysRevE.81.011109}. Coarse-grained observables need not be 
 the result of only local measurements. In appendix \ref{coarse} we briefly summarize the current understanding of
  equilibration for coarse-grained observables.
%


\section{Intuition from finite systems}

Recent developements in quantum information theory have made it possible to reformulate the foundations of thermodynamics in a more
 precise way. The authors in \cite{citeulike:3788627,reimann2010canonical,citeulike:9264197}, among many others, revisited the question of equilibration
  in the quantum world and proved the following three statements about weak equilibration in finite-dimensional systems that we refer to as
  fluctuation, typicality and universality theorems:

\begin{theorem}{\bf{Fluctuations:}}
Consider an arbitrary initial state $\psi_0$ in a finite dimensional Hilbert space $\cH_R\subset\cH$ evolving with a Hamiltonian with non-degenerate energy gaps.\footnote{For any equilibration to happen in Hilbert space it is essential that 
the degrees of freedom interact.
One way to assure this is to restrict the problem to Hamiltonians with no energy gap $E_i-E_j$ that is hugely degenerate. In a free
 theory $E=E_S+E_{\bS}$, so for any arbitrary $E_{\bS}$ and energy gap $\delta E_S$in $S$, there is a gap $\delta E$
 in $E$. 
  Note that even 
 an arbitrarily small generic interaction lifts all degeneracies. 
 The generalization of this result to the case of degenerate interacting Hamiltonians
is discussed in \cite{citeulike:11048022}.}  
Let $\omega=\la \psi(t)\ra_t$ be the time-averaged density matrix of the whole system, and 
$d_{eff}(\omega)=\frac{1}{\tr(\omega^2)}$ be a measure of the effective dimension of the Hilbert space it explores.
For any subsystem $S$ we define:
\bea
&&\eta_S(\psi_0)=\sqrt{\frac{d_S^2}{d_{eff}(\omega)}},\nn
\eea
where $d_S$ is the dimension of $\rho_S$.
Then,
\bea
\la \|\rho_S(t,\psi_0)-\omega_S(\psi_0)\|\ra_t\leq \eta_S(\psi_0).
\eea
Therefore, for all initial states $\psi_0$, $\rho_S(t,\psi_0)$ approaches its time average in a weak sence if $\eta_S(\psi_0)\ll 1$. 
  
\end{theorem}

Hereafter, we refer to $\eta_S$ as the equilibration parameter.
Roughly speaking, $\eta_S$ bounds the size of flucutations around equilibrium. we expect $\eta_S$ to be small whenever the subsystem $S$ is much smaller than the system.
One way to make this intuition more percise is to ask: what is the probability
of finding a large $d_{eff}(\omega)$ if we pick a random $\psi_0$ from $\cH_R$? The answer to this question is
given by the typicality theorem which makes use of Levy's lemma \footnote{For completeness, we have included Levy's lemma 
 in appendix \ref{levy}.}:

\begin{theorem}{\bf{Typicality:}}
 For a state $\psi_0$ chosen at random with uniform measure in $\cH_R$:
 \bea
  \text{Prob}\left(d_{eff}(\omega)<\frac{d_R}{4}\right)\leq 2 e^{-c\sqrt{d_R}},
 \eea
where $c=\frac{(\ln 2)^2}{72\pi^3}\approx 2\times 10^{-4}$.
\end{theorem}

Putting the first two theorems together we conclude that as long as the dimension of
the initial ensemble's Hilbert space $\cH_R$ is much larger than the dimension of 
a subsystem's Hilbert space, any interacting Hamiltonian evolves the subsystem towards equilibration. Naturally, 
we expects the equilibrium state to be almost independent of the initial state $\psi_0$. The universality
 theorem quantifies this by showing that the equilibrium states corresponding to all states in $\cH_R$ are almost indistinguishable:
 
 \begin{theorem}{\bf Universality:}
 For a random state $\psi_0$ chosen with uniform measure in $\cH_R$:
 \bea
 \text{Prob}\left(\|(\omega(\psi_0),\la\omega\ra_{\psi_0}\|>\frac{1}{2}\sqrt{\frac{d_S}{d_R}}+\ep\right)\leq 2 e^{-c'\ep^2 d_R},
 \eea
 where $c'=\frac{2}{9\pi^3}$. Here $\la\omega\ra_{\psi_0}$ is the state $\omega(\psi_0)$ averaged over $\cH_R$.\footnote{The tighter version of this bound
  introduced in \cite{citeulike:3788627} is crucial for our discussion of scrambling in section 5. However, this form suffices for our purposes here.}\\
   \end{theorem}
Figure 2 summarizes the approach discussed above.
\begin{figure}\label{fig2}
\begin{center}
\input{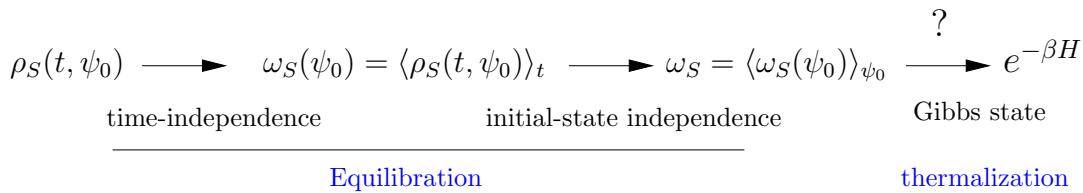}
\caption{Information-theoretic approach to equilibration: First, the fluctuation and typicality theorems establish the time-independence of 
the subsystem density matrix by showing that $\rho_S(t)$ is close to its time-average for a typical state. Then, the time-averaged state is shown to be close to its initial ensemble average.}
  \end{center}
\end{figure}

\section{Equilibration in infinite-dimensional Hilbert spaces}
In systems with bosonic degrees of freedom, $\rho_S$ lives in an infinite dimensional Hilbert space. 
Equilibration in trace norm is the condition that
 $\rho_S$ at any time is indistinguishable from its equilibrium value with respect to an infinite 
 number of independent measurements. It is not apriori clear that such a strong condition holds in equilibrated field theories. 
 This problem is manifest in the appearance of $d_S$ in the equilibration parameter. Subsystems can have infinite dimensions, and since 
 the denominator in $\eta_S$ only depends on the initial state and is finite when $d_R$ is finite, the bound seems useless.
 
 What saves us is the intuition that in any physical equilibration process the support of the wave function on 
arbitrarily large energy eigenstates is small. Therefore, one should be able to truncate the Hilbert space by reasonable 
dynamical assumptions without missing out the relevant physics. As long as the errors caused by truncation are small, the finite dimensional 
calculations are reliable. When the equilibration parameter is small but truncation errors are not, it is 
 only the set of observables which act on the truncated subspace that equilibrate. Significant deviations from
 equilibrium can be detected using operators that act beyond the truncated space.

  Cutting off the full Hilbert space in the interacting Hamiltonian basis at energy $\Lambda$
does not guarantee that the reduced density matrix on $S$ is finite-dimensional. 
However, if we instead truncate in the free Hamiltonian basis, 
 not only does it regulate $d_S$ but also it allows us to estimate the equilibration parameter 
by counting partitions.
 
In order to make the discussion more concrete, we choose to study the following ``quench'' problem.\footnote{Typically the term quench is used when the initial state is the
ground state of an initial Hamiltonian. Here, the initial state is chosen from some large energy sector of the free theory.}
Consider a free theory on $\mf=\mf_1\otimes...\otimes\mf_N$ with Hamiltonian 
  $H_{free}=\sum_{i=1}^N\mu_iH_i$. The Hilbert space is a direct sum $\cH=\oplus_E\cH_E$, where $\cH_E$ is 
  the eigenspace corresponding to energy $E$ spanned by $e_E=\{|n_1,\hdots,n_N\ra:\:E=\sum_i\mu_in_i\}$. We choose our
   initial states from a subspace corresponding to energy $\Delta$: $\psi_0\in\cH_R=\cH_\Delta$.\footnote{Choosing $\psi_0\in \oplus_{E=\Delta-\delta}^{\Delta+\delta}\cH_E$ with some small $\delta/\Delta$ 
   will not change any of our general conclusions.} Initially $d_R$ is the cardinality of $e_\Delta$, and $d_S$ is 
   the number of distinct $|n_{k_1},\hdots,n_{k_s}\ra$ one finds in $e_\Delta$, by restricting to subsystem's degrees of freedom: $S=\{k_1,\hdots,k_s\}$.
   
   At $t=0$ we quench the system by turning on an interaction Hamiltonian, and study the relaxation of $\rho_S(t)$. 
   If we assume that $[H_{int},H_{free}]=0$, it is clear that $\psi(t)$ never leaves $\cH_\Delta$ ,and the problem remains finite-dimensional.
   Hence, the results of previous section apply: subsystem $S$ equilibrates if $d_S^2\ll d_\Delta$.  
  Generically $[H_{int},H_{free}]\neq 0$, and both $\psi$ and $\rho_S$ have infinite-dimensional Hilbert spaces available to them to explore. Since $d_\Delta$ only depends on the initial ensemble 
  it remains finite, whereas $d_S$ is unbounded above and $\eta_S$ diverges. This is
  an artifact of the infinite dimensionality of the Hilbert space.
  Let us define the projector to the finite-dimensional Hilbert space $\oplus_{E<\Lambda} \cH_E$:
   \bea\label{proj}
 P^\Lambda|\psi\ra=\sum_{\sum_i \mu_in_i=0}^{\Lambda}|n_1,\hdots,n_N\ra\la n_1,\hdots,n_N|\psi\ra.
 \eea
 In a similar fashion, $P^\Delta$ denotes the projector to the initial ensemble $\cH_\Delta$.
 We expect that there exists a truncation such that $\rho_S^\Lambda(t)=\tr_{\bS}(P^\Lambda\psi(t)P^\L)$ approximates $\rho_S(t)$ well. Here $\bar{S}$ denotes the complement of $S$.    
%
    The triangle inequality tells us:
    \bea\label{trunc}
  \|\rho_S(t)-\omega_S\|&&\leq \|\rho_S^\Lambda(t)-\omega_S^\Lambda\|+\|(\rho_S(t)-\omega_S)-(\rho_S^\Lambda(t)-\omega_S^\Lambda)\|
 \eea
  We call the second term on the right hand side the truncation error, and denote it by $\mathcal{E}^\L$.
  The trick is to choose a cut-off $\Lambda$ such that both the equilibration parameter in the truncated subspace $\eta_S^\Lambda$, and the errors caused by truncation 
  are small.
  For this procedure to work, we need to make 
  a dynamical assumption that limits the support of wave-function on $\cH_E$ for $E>\Lambda$, and
  quantifies the truncation errors. Here, we employ a generalization of the method in
  \cite{ohliger2011continuous} that uses the mean photon number in quantum optics to 
  quantify errors caused by truncation to a finite-dimensional Hilbert space. 
  The following theorem provides upper bounds on both terms on the right hand side of (\ref{trunc}):

\begin{theorem}{\bf{Equilibration in infinite-dimensional $\cH$:}}\label{theo}
Consider an arbitrary state $\psi_0$, chosen at random from a subspace corresponding to energy $\Delta$ in the Hilbert spcae of a free theory. 
For any subsystem $S$, and projectors $P^\Lambda$ and $P^\Delta$ defined as in (\ref{proj}), we define:
\bea
&&\eta^\Lambda_S(\psi_0)=\sqrt{\frac{(d_S^\Lambda)^2}{d_{eff}(\omega)}},\nn
\eea
where $d_S^\Lambda$ is the dimension of the projected Hilbert space available to $S$. After a quench by 
an arbitrary interaction Hamiltonian:
  \begin{enumerate}[(i)]
\item {\bf Fluctuations:}
\bea
\la \|\rho_S(t)^\Lambda-\omega_S^\Lambda\|\ra_t\leq \eta^\Lambda_S.
\eea
\item {\bf Truncation error:}
\bea\label{trunc}
\mathcal{E}^\L\equiv\la\|(\rho_S(t)-\omega_S)-(\rho_S^\Lambda(t)-\omega_S^\Lambda)\|\ra_t\leq 6\sqrt{\frac{\tr(\omega H_{free})}{\Lambda}}
\eea
 
 \item {\bf Typicality:}
 \bea
  \text{Prob}\left(d_{eff}(\omega)<\frac{d_\Delta}{4}\right)\leq 2 e^{-c\sqrt{d_\Delta}},
 \eea
where $c=\frac{(\ln 2)^2}{72\pi^3}\approx 2\times 10^{-4}$.

 \item {\bf Universality:}
 \bea\label{uni}
 &&\text{Prob}\left(\|\omega^\Lambda_S(\psi_0)-\la\omega^\Lambda_S\ra_{\psi_0}\|>\frac{1}{2}\sqrt{\frac{d_S^\L\delta^\L}{d_\Delta}}+\ep\right)\leq 2 e^{-c'\ep^2 d_\Delta},\\
  &&\delta^\L=\sum_k\la k| \frac{P^\Delta}{d_\Delta}|k\ra \:\tr[\tr_{\bS}(P^\L|k\ra\la k|P^\L)^2],
 \eea
 where $c'=\frac{2}{9\pi^3}$. Here $\la\omega\ra_{\psi_0}$ is the state $\omega^\Lambda(\psi_0)$ averaged over $\cH_\Delta$, and 
 $|k\ra$ are energy eigenstate of the interacting Hamiltonian. 
        \end{enumerate}
  \end{theorem}

 The above theorem implies that the state of subsystem $S$ equilibrates for almost all initial states $\psi_0\in\cH_\Delta$, if there exist a cut-off $\Lambda$ such that $\eta^\Lambda_S\ll 1$ and $\tr(\omega H_{free})\ll\Lambda$.
The proof appears in appendix \ref{appC}.

The dynamical condition to meet to justify a truncation 
with $P^\Lambda$ is that the time-averaged expectation value of the free part of the Hamiltonian, which was initially $\Delta$, is much smaller than the cut-off scale $\Lambda$.\footnote{Roughly speaking, this is a constraint on the number density of particles.} 
For systems with $N$ degrees of freedom, it is more convenient to define new variables $p$, $m$, $\bar{m}$ and $\lambda$:
\bea
&&m=\Delta/N^p\nn\\
&&\bar{m}=\tr(\omega H_{free})/N^p\nn\\
&&\l=\L/\Delta,
\eea
that remain finite in the thermodynamic limit $N\to\infty$.
In terms of these new variables, the truncation error $\mathcal{E}^\l\leq 6\sqrt{\frac{\bar{m}}{m\l}}$.
We devote the rest of this section to discussing the implications of theorem \ref{theo}
in three different physical systems.

\subsection{Bosons on a lattice}
As our first example of equilibration in multi-particle quantum systems we consider bosons on a 
lattice with $N$ sites. The Hamiltonian of the free theory, in appropriate units, is simply the total boson number:
\bea
H=\mu N=\mu\sum_i\:a_i^\dagger a_i,
\eea
 where $a_i$ and $a_i^\dagger$ are the creation and annihilation operators defined at each site on the lattice.
 Setting $\mu=1$, the energy spectrum is given by all non-negative integers $\Delta$, each of which appear with 
degeneracy $d_\Delta={\Delta+N-1 \choose N-1}$, which is the number of weak compositions of $\Delta$ in $N$ parts.{\footnote{A weak composition of 
$\Delta$ in $N$ parts is a sequence of $N$ non-negative integers which sum to $\Delta$.}  
We restict the ensemble of initial states to $\cH_\Delta$. The effective dimension of $\psi_0$ only depends on the initial ensemble, and for almost all $\psi_0$ is larger than $d_\Delta/4$. 
 Whereas, the dimension of the subspace available to subsystem $S$ after introducing the cut-off $\Lambda$ is the number of distinct compositions one finds by 
  restrciting to a sequence of length $s=|S|$ in weak compositions of $\Lambda$ in $N$ parts:
  \bea\label{dslam}
  d_S^\Lambda\leq\sum_{i=1}^{\Lambda-1}{i+s-1\choose s-1}=\frac{\Lambda}{s}{\Lambda+s-1\choose s-1}-1
  \eea
  We are interested in investigating how subsystems of different sizes equilibrate in the thermodynamic limit $N\to\infty$:

  {\bf Finite-site subystems:}
  The density matrix of subsystems of size $s=O(1)$ contains information about small observables of the sort $\la \phi^{m_1}(x_1)\hdots\phi^{m_s}(x_s)\ra$.
 Let us assume for the moment that the number of bosons per site is initially finite and remains so at all times, i.e. $p\leq1$.
    From (\ref{trunc}) we find that the error caused by truncation by $\l=N^\ep$ 
   is negligible for any $\ep>0$. 
   Plugging these into the expression for the equilibration parameter gives
   \bea
   \eta_S^\L<\frac{N^\ep\Delta}{s!}\sqrt{\frac{(N^\ep\Delta+s-1)!^2(N-1)!\Delta !}{(N^\ep\Delta)!^2(\Delta+N-1)!}}
   \eea
  Using the Stirling approximation\footnote{$1\leq \frac{n!}{\sqrt{2\pi n}(n/e)^n}\leq \frac{e}{2\pi}$.} we find that:
   \bea
   \eta_S^\Lambda\leq\left\{ 
  \begin{array}{l l}
    f(s)\:N^{s(p+\ep)+p/4-(1-p)m N^p/2} & \quad \text{if $0<p<1$}\\
    g(s)\:N^{(1+\ep)s+1/4}\:\left(\frac{m}{1+m}\right)^{m N/2} & \quad \text{if $p=1$},
  \end{array} \right.
   \eea
   for some $O(1)$ functions $f(s)$ and $g(s)$.
Hence, in the thermodynamic limit $N\gg 1$ the density matrix of all finite subsystems equilibrate.
%

 {\bf A small but finite fraction of all sites:} Consider subsystems of size $s/N$ order one but small, 
with a finite initial boson number density per site: $m=O(1)$. 
Applying the Stirling approximation one finds that the equilibration parameter is bounded by
\bea
\eta^\lambda_S\leq&&c \left(\left(\frac{m N \lambda\:\sqrt{e}}{s}\right)^{2s/N}\frac{1}{(1+m)}\right)^{N/2}\:\left(\frac{m}{1+m}\right)^{m N/2}\:\left(\frac{N}{s^2}\right)^{1/4}\nn\\
<&&c \left(\left(\frac{m N \lambda\:\sqrt{e}}{s}\right)^{2s/N}\frac{1}{(1+m)}\right)^{N/2},
\eea
for some $O(1)$ constant $c$.
For any initial energy per site $m$ and small error $\mathcal{E}^\l$, there exists an $s_0$ found from 
\bea\label{alpha0}
\left(\frac{m N \lambda\:\sqrt{e}}{s_0}\right)^{2s_0/N}=(1+m)
\eea
such that any subsystem smaller than $s_0$ equilibrates. 

%
%
%

{\bf Large subsystems:}   
A careful look at equation (\ref{alpha0}) shows that for any fixed $\lambda$, $s_0(m)$ is an increasing function of $m$ with the asymptotic value $\lim_{m\to\infty}s_0=N/2$; see figure \ref{figure1}
. 
  It is a curious fact that for large enough initial boson number density, not only small subsystems  
   but also all subsystems smaller than half the size of the system equilibrate.
%
   Intuitively, this is due to the exponential growth in density of states as a function of energy which makes the fraction of the Hilbert space
 available to $S$ vanishingly small for $N\gg 1$. In fact, for any $\Delta>O(N)$, in the limit of $N\to\infty$, both the error and 
 the equilibration parameter vanish for any subsystem smaller than half.
\begin{figure}\label{figure1}
\begin{center}
 \includegraphics{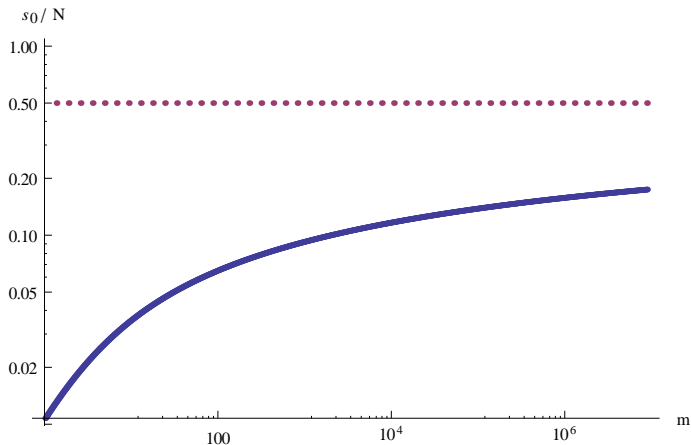}
 \caption{The size of the largest subsystem that provably equilibrates plotted as a function of the initial average boson number per site. The error parameter is taken to be $\mathcal{E}^\lambda\leq 6\times10^{-6}\sqrt{\bar{m}/m}$. In the limit of infinite initial energy 
 any subsystem smaller than half the size of the whole system equilibrates.}
 \end{center}
\end{figure}

  %
%
%
%


\subsection{Hermitian matrix model}
For bosons on a lattice we found that states with energies $\Delta>O(N)$ equilibrate
  on all subsystems smaller than half. However, we do not expect such a sector to survive in continuum field theories.
 An example where one is interested in studying such high energy states is the matrix model. 
 Consider the Hermitian one-matrix model with a free Hamiltonian
 \bea\label{Hmatrix}
 H=\tr\left(\dot{X}^2+m^2 X^2\right),
 \eea
 where $X$ is an $N\times N$ matrix. The Hamiltonian is $U(N)$ invariant, a freedom which can be partially
 fixed by diagnalizing the matrix. In the diagnal form, each eigenvalue $\lambda_i$ becomes a simple harmonic oscillator. 
 The change of variables in the quantum partition function gives a Van der Mond determinant which can be absorbed by a redefinition of 
 eigenvalues \cite{brezin1978planar, Berenstein:2004kk}. In terms of redefined eigenvalues, the wave-function $\psi(\lambda_1,\hdots,\lambda_N)$ is a tensor product of
  $N$ harmonic oscillators states, anti-symmetrized under the permutation group of eigenvalues:
\bea
\psi(n_1,...,n_N)=\frac{1}{\sqrt{N!}}\:\det
 \begin{pmatrix}
  |n_1(\l_1)\rangle & |n_1(\l_2)\rangle & \cdots & |n_1(\l_N)\rangle \\
  |n_2(\l_1)\rangle & |n_2(\l_2)\rangle  & \cdots & |n_2(\l_N)\rangle \\
  \vdots  & \vdots  & \ddots & \vdots  \\
  |n_N(\l_1)\rangle & |n_N(\l_2)\rangle  & \cdots & |n_N(\l_N)\rangle
 \end{pmatrix},
\eea
where $|n_i(\l_j)\rangle$ is the $n_i^{th}$ excited state of the $j^{th}$ eigenvalue and multiplication inside the determinant is tensor product of states. The fermionic nature of the degrees of freedom implies that
$n_i$ is ordered as $n_N>n_{N-1}>\hdots>n_1$. The ground state has energy
\be
E^{(0)}=\sum_{k=0}^{N-1}\left(k+\frac{1}{2}\right)\omega=\frac{N^2}{2}\omega.
\ee
We will set $\omega=1$ and measure energy in units of $\omega$. The excitations are described by an ordered set of positive numbers
$r_N\geq r_{N-1}\geq \hdots r_1\geq 0$ where each $r_i$ corresponds to the number of times the $i^{th}$ 
energy level is excited over its ground state, i.e.
\be
r_i=E_i-E_i^{(0)}.
\ee
 If $\Delta=E-E^{(0)}$, then to each total energy $E$ corresponds a sub-Hilbert space $\cH_\Delta$ spanned by degenerate eigenstates of that energy level. 
The dimension of $\cH_\Delta$ is $P(\Delta,N)$, the number of paritions of $\Delta$ into at most $N$ parts. To each of these states one can
associate a Young diagrams with $r_{N-i+1}$ boxes in the $i^{th}$ row from the top \cite{Berenstein:2004kk,Balasubramanian:2005mg}. Then, $P(\Delta,N)$ counts the 
number of Young diagrams with $\Delta$ boxes and maximum $N$ rows.

Consider a subsystem of $s$ eigenvalues. The dimension of $d_S^\Lambda$ is bounded above by the number of distinct subdiagrams one finds by
 picking $s$ rows from Young diagrams corresponding to $P(\Lambda,N)$, and discarding the rest of the rows. Sub-diagrams have 
 any number of boxes from one to $\L$, but have at most $s$ rows. Therefore,
 \bea
 d_S^\Lambda\leq \sum_{E=1}^\Lambda P(E,s)<\Lambda\: P(\Lambda,s).
 \eea
 Then, the equilibration parameter satisfies
 \bea
 \eta^\Lambda_S\leq \Lambda\sqrt{\frac{P(\Lambda,s)^2}{P(\Delta,N)}}.
 \eea
By studying the large $N$ asymptotics of restricted partitions functions $P(N,s)$ we find\footnote{See appendix \ref{asymp}.}:

  {\bf A finite set of eigenvalues:} For $s=O(1)$ and $p\geq 0$, the equilibration parameter is small:
  \bea
  \eta^\Lambda_S\leq  c\Lambda^s e^{-1/2 \left(N^{1/2\min[p,1]}\right)}
  \eea
Choosing $\Lambda$ to be 
 any polynomial of finite order larger than one in $\Delta$ keeps both the truncation error and the equilibration parameter small.
  Hence, in the thermodynamic limit, small subsystems equilibrate.
  
  {\bf Large subsystems:} One particularly interesting regime is $\Delta=O(N^2)$. 
In this regime, we find that for large $N$ and $s/N=O(1)$, the equilibration parameter is bounded by
 \bea
 \eta_S^\Lambda\leq c\:\sqrt{N}\: e^{f(\alpha,m,\lambda)N/2},
 \eea
 for an order one constant $c$. The exponent on the right hand side is given by:
 \bea\label{alpha0matrix}
 f(s,m,\lambda)&&=2\left(\beta_c \frac{s}{N}-\frac{m N\lambda }{s}\log(1-e^{-\beta_c})\right)-\left(\beta-2m\log(1-e^{-\beta})\right),\nn\\
 \eea
 where $\beta$ and $\beta_c$ are found by solving
 \bea
 &&m=\frac{Li_2(e^{-\beta})}{\log(1-e^{-\beta})^2}\nn\\
 &&\frac{mN^2\lambda}{s^2}=\frac{Li_2(e^{-\beta_c})}{\log(1-e^{-\beta_c})^2}.
 \eea
  \begin{figure}
\begin{center}
 \includegraphics{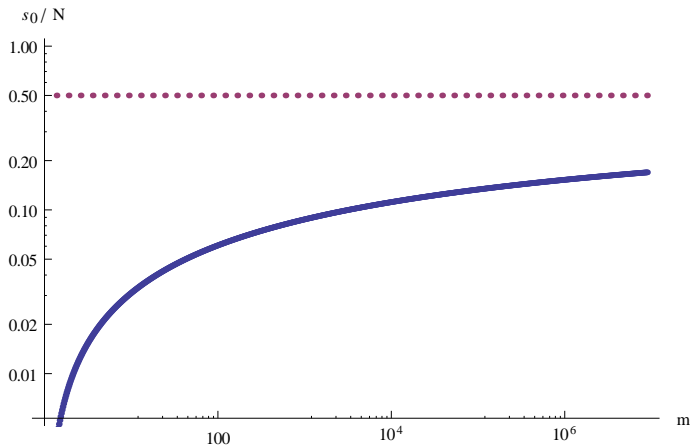}
 \caption{The size of the largest subsystem that provably equilibrates plotted as a function of the average initial energy per matrix element; $\Delta/N^2$. The error parameter is suppressed by $\mathcal{E}^\lambda\leq 6\times10^{-6}\sqrt{\bar{m}/m}$. In the limit of infinite initial energy 
 any subsystem smaller than half the size of the whole system equilibrates.}
 \end{center}
\end{figure}
 For fixed $m$ and arbitrary small error $\mathcal{E}^\lambda\leq 6\sqrt{\bar{m}/(m \lambda)}$ all subsystems of size smaller than $s_0 N$ equilibrate, where $s_0$ is the solution to $f(s_0,m,\lambda)=0$. 
 In the limit of a large initial boson number $m\gg 1$, and small errors $\mathcal{E}^\lambda\ll 1$, it is possible to simplify the expressions in (\ref{alpha0matrix}) 
 to obtain:
 \bea
 m\simeq\left(\frac{N^2\lambda}{s_0^2}\right)^{2s_0/(N-2s_0)}
 \eea
  The interesting fact is that by increasing $m$ it becomes possible for larger and larger subsystems to equilibrate. In the energy range $\Delta>O(N^2)$ 
  we find that all subsystems smaller than half the size of the system equilibrate with a vanishing error. Note that in the Hermitian matrix model, the regime relevant for equilibration
  of large subsystems has an initial energy per degree of freedom that scales with $N$, as opposed to bosons on a lattice where it was finite.

\subsection{Field theories in momentum space}

Our discussion so far has been restricted to discrete models. 
In continuum field theories the Hilbert space of free theory has a natural decomposition 
in terms of momentum modes.
 On a compact space and in the absence of any internal symmetries, the Hilbert space is a tensor product of infinite Fock spaces
$\mf_k$ corresponding to momentum modes $k$. There is no analogue of parameter $N$ in momentum space, so we only consider high energy ensembles, i.e. $\Delta\gg 1$.

\subsubsection*{1+1-dimensional field theories}

For simplicity, let us start with a scalar field 
in a one-dimenstional box and normalize such that $k$'s are positive integers.\footnote{The generalization to fields with non-zero spin is straightforward.} The Hamiltonian of a free massless theory has the form
\bea
H=\sum_{k}^\infty k\:a^\dagger_ka_k.
\eea
The dimension of the energy eigenspace $\Delta$ is given by the number of partitions of $\Delta$, which according to the Hardy-Ramanujan expansion scales as
\bea
d_\Delta=P(\Delta)\simeq e^{\pi \sqrt{2\Delta/3}}.
\eea
On the other hand, the dimension of the Hilbert space available for a set of momentum modes $S=\{k_1,\hdots,k_s\}$ with a momentum cut-off $\Lambda$ satisfies
  \bea
  d_S^\Lambda\leq\sum_{E=1}^\Lambda D(E;S),
  \eea
 and $D(E;S)$ is the denumerant; the number of ways one can partition the number $E$ as $a_1k_1+\hdots+a_sk_s$.
 
 {\bf Finite momentum modes:}
 The large $E$ asymptotic of denumerant for finite $s$ is given by\footnote{See appendix \ref{asymp} for details.}
    \bea\label{denum}
  D(E;S)\simeq\frac{E^{s-1}}{(s-1)!\:k_1k_2\hdots k_s}.
  \eea
   It is clear from (\ref{denum}) that for large $\Lambda$, $d_S^\Lambda$ is bounded by a polynomial of degree $s$ in $\Lambda$. As a result,
    for 
   any $\Lambda$ polynomial in $\Delta$ of a finite order larger than one, both the equilibration parameter 
    and the truncation error are small.
   The theorem \ref{theo} implies that $\rho_S$ equilibrates for any finite set of momentum modes.
   In other words, the results of all field theory measurements and observables that involve a 
   finite number of $k$ modes, almost at all times after the quench, are well approximated by 
those of a stationary state, conditioned that whatever the non-perturbative dynamics is, the expectation value of kinetic
energy does not grow indefinitely in time.
 
%
%
{\bf Large subsystems:} An interesting observation is that
   for any subsystem size $s$, $d_S^\Lambda$ is at its maximum value when $S$ is the ``infra-red'' density matrix; 
   i.e. $S=\{k:|k|\leq s\}$.\footnote{Note that in a theory with non-perturbative
    interactions, low-momentum modes can have overlap with large energy eigenstates. Therefore, the reduced density matrix $\rho_S$ is not necessarily capturing the low-energy dynamics.
     However, the constraint on the time-average of the free Hamiltonian expectation value implies that this overlap is small for eigenstates of energy larger than $\Lambda$.} 
     The reason is that for all $S$ and $E$ the denumerant satisfies $D(E;S)\leq P(E,s)$, where $P(E,s)$ is the number of 
   partitions of $E$ into at most $s$ parts with the equality holding when
   $S=\{k:|k|\leq s\}$. \footnote{We are using the following well-known result in number theory:  
   the number of partitions of $E$ into at most $m$ parts is equal to the number of partitions of $E$ with each part at most $m$.}
  
 Let $\rho^{IR}_\mu$ and $\rho^{UV}_\mu$ denote, respectively, the ``infra-red'' and ``ultra-violet'' density matrices corresponding to the set of modes $S=\{k:k\leq\mu\}$ and $S=\{k:k>\mu\}$.
   For a finite energy state, the expecation is that we do not lose  
  details of the dynamics by tracing out large momentum modes. In this context of equilibration, we ask what are the largest and smallest) $\mu$ such that, respectively,  
  $\rho^{IR}_\mu$ and $\rho^{UV}_\mu$ are time-independent, and contains no information about the dynamics or the initial state.
    
  The dimension of the Hilbert space available to $\rho^{IR}_\mu$ is 
  \bea
  d_\mu^{IR}\leq\sum_{E=1}^\Lambda P(E,\mu).
  \eea
  The asymptotic behaviour of $P(E,\mu)$ for all large $\mu$ and $\Delta$ is such that the equilibration parameter is
   never small unless $\mu=a \sqrt{\Delta}$, for some $a$ at least order one.
  The analysis of asymptotics of restricted partitions in appendix \ref{asymp} shows that for $\mu=a \sqrt{\Delta}$:
   \bea
       &&\eta_S^\Lambda\leq c\:\sqrt{\Delta}\:e^{\frac{1}{2}g(a)\sqrt{\Delta}},
       \eea
       where
       \bea\label{momen}
        &&g(a)=a\left(\frac{2}{\tbeta}\int_0^{\tbeta}\frac{dt\:t}{e^t-1}-\log(1-e^{-\tbeta})\right)-\sqrt{\frac{2}{3}}\pi\nn\\
          &&\tbeta^2=\frac{a^2}{\lambda}\int_0^{\tbeta} \frac{dt\:t}{e^t-1},
   \eea
  and $c$ is some order one constant.
   For $\lambda$ large enough so that $\mathcal{E}^\l\ll1$, the equilibration parameter is small for all $a\leq a_{IR}$, where $a_{IR}$ is the solution to
   $g(a_{IR})=0$.
  
  Similarly, the dimension of the Hilbert space available to $\rho^{UV}_\mu$ is
  \bea
  d_\mu^{UV}=\sum_{E=1}^\Lambda P(E,\geq \mu),
  \eea
  where $P(E,\geq \mu)$ is the number of partitions of $E$ into parts at least $\mu$. Since these partitions have at most $E/\mu$ parts, then $P(E,\geq\mu)\leq P(E,E/\mu)$.
  It follows that $\rho^{UV}_\mu$ equilibrates if $a> a_{UV}$ where $a_{UV}$ is found by replacing $a$ by $\lambda/a$ in (\ref{momen}) and solving for $g(a_{UV})=0$.
%
%
%

  \subsubsection*{Higher-dimensional field theories}\label{higher}
  
   In dimensions higher than one, it is easier to work with generating partition functions:
 \bea
   Z(q)=\sum_{\Delta}P(\Delta)q^\Delta=\prod_{k_{x_1},\hdots,k_{x_d}}\frac{1}{(1-q^{\Delta(k)})}.
   \eea
   For simplicity we assume that the field theory lives in a cubic box.\footnote{Knowledge of the spectrum of Laplacian is sufficient to generalize this method to arbitrary compact spaces.}
   Then, the free energy at large $\Delta$ is approximated by an integral:
   \bea
   F(e^{-\beta})=&&-\frac{1}{\beta}\log Z(e^{-\beta})=\Omega(d-1)\beta^{-1}\:\int_0^\infty dk\:k^{d-1}\:\log(1-e^{-\beta k})\nn\\
   &&=-\Omega(d-1)\beta^{-1}\sum_{n=1}^\infty\int_0^\infty dk\:k^{d-1}\:\frac{e^{-\beta nk}}{n}\nn\\
   &&=-\beta^{-(d+1)}f(d),
   \eea
   where $f(d)=\Omega(d-1)\Gamma(d)\zeta(d+1)$.
   The expectation value of energy and entropy are found in terms of $\beta$: 
   \bea
   &&\bar{E}=F+\beta\p_\beta F=\beta^{-(d+1)}\:d\:f(d)\nn\\
   &&S(\beta)=\beta^2\p_\beta F=\beta^{-d}\:(d+1)\:f(d).
  \eea
  The number of partitions at large $E$ is well approximated by $e^{S(\bar{E})}$:
  \bea\label{ent}
  P(E)\simeq e^{A E^{d/(d+1)}}
  \eea
    where $A=(d+1)\:(d^{-d}\:f(d))^{1/(d+1)}$. This is a well-known result for density of states in free $d$-dimensional theories.
  
  The partition function that generates $P(E,s)$ is also easy to write down:
  \bea
  Z_S(\beta)=\sum_E P(E,s) e^{-\beta E}=\prod_{k_{x_1},\hdots,k_{x_d}}^{|k|=s}\frac{1}{(1-e^{-\beta E(k)})}
  \eea
  The free energy associated with this partition function is 
\bea
F&&=\Omega(d-1)\beta^{-1}\:\sum_{n=1}^\infty \int_0^{s} dk\:k^{d-1}\:\log(1-e^{-\beta k}).
\eea
The expectation value of energy and the entropy are
\bea\label{expE}
\bar{E}&&=\Omega(d-1)\int_0^{s}\frac{dk \:k^d}{e^{k\beta}-1}=\Omega(d-1)\tilde{\beta}^{-(d+1)}s^{(d+1)}\int_0^{\tilde{\beta}}
\frac{dt\: t^d}{e^t-1}\\\label{expS}
S(\beta)&&=\Omega(d-1)\left(\beta\int_0^{s}\frac{dk\:k^d}{e^{\beta k}-1}-\int_0^{s}dk\:k^{d-1}\:\log(1-e^{-\beta k})\right)\nn\\
&&=\Omega(d-1)s^d\left(2\tbeta^{-d}\int_0^{\tbeta}\frac{dt\:t^d}{e^t-1}-\frac{\log(1-e^{-\tbeta})}{d}\right)
,
\eea
where we have defined $t=\beta k$ and $\tbeta=\beta s$.\footnote{For $d=1$, after some algebra \ref{expE} reproduces the asymptotic behaviour we quote in appendix \ref{asymp}.}
When $\bar{E}\gg 1$ and $\frac{s}{E^{1/(d+1)}}\gg 1$, (\ref{expE}) implies that $\tbeta$ is small.
The entropy at small $\tbeta$ is dominated by the term $(s/\tbeta)^d$ in (\ref{expS}). Solving for entropy as a function 
of energy in this limit reproduces the familiar expression in (\ref{ent}) for entropy of free theories. In this regime, since
 $d_S^\L\simeq d_\Delta$, the equilibration parameter is large. However, if $\frac{s}{E^{-(d+1)}}=O(1)$ then
  $\tbeta$ is also order one, and $P(E,s)$ grows as $E^{d/(d+1)}$, but with a smaller coefficient. Similar to the situation in one spatial dimension,
 any subsystem of size $s=a E^{1/(d+1)}$ with $a<a_{IR}$ equilibrates, where $a_{IR}$ is the solution to $g(a_{IR})=0$ for
 \bea
 &&g(a)=\Omega(d-1)a^d\left(2\tbeta^{-d}\int_0^{\tbeta}\frac{dt\:t^d}{e^t-1}-\frac{\log(1-e^{-\tbeta})}{d}\right)-\frac{A}{2}\nn\\
 &&a^{-(d+1)}=\Omega(d-1)\frac{\tbeta^{-(d+1)}}{\lambda}\int_0^{\tbeta} \frac{dt\:t^d}{e^t-1}.
  \eea



\section{Atypical initial states and scrambling}

The typicality theorem implies that almost all initial states (typical states) in $\cH_R$ equilibrate; however,  
in scrambling we are interested in far from equilibrium initial states which are highly atypical.
For instance, consider an initial product state $\psi=\phi_M\otimes \psi_{\bar{M}}$ for a small subsystem $M$. Then, from
the fluctuation theorem we know that for a fixed $\phi_M\in\cH_M$, and typical states $\psi_{\bar{M}}$ 
 \bea
 \la\|\rho_S(t)-\omega_S(\psi_{\bar{M}},\phi_M)\|\ra<\sqrt{\frac{d_S^2}{d_{eff}(\omega_{\bar{M}})}}.
 \eea
As a result, if $d_{\bar{M}}\gg d_S^2$ the density matrix of $S$ is independent of $\psi_{\bar{M}}$. 
As we saw previously, this condition is easily satisfied for small messages.\footnote{
Replacing $d_R$ with $d_R^{\bS}$ in theorem \ref{theo} implies typicality and universality for product states with typical $\psi\in\cH_R^{\bS}$ .} 
However, the equilibrium density matrix can still depend on the message $\phi_M$. We say a Hamiltonian 
scrambles an ensemble of initial states, if all subsystems $S$ smaller than half the size of the system $\omega_S$ become independent of
 $\phi_M$. It is clear that this does not always occur. As was argued in \cite{citeulike:3788627}, in systems where the eigenstates of 
 the interacting Hamiltonian remain close to tensor products on $M$ and $\bar{M}$ (as in some weakly coupled systems), $\omega_S$ retains 
 information about $\phi_M$. 
 
One might think that the type of arguments we have been using so far fails to provide insight 
 about scrambling of small messages. The typicality arguments we have been using are based on the idea  that
  a small subsystem $S$ does not have enough Hilbert space (memory) available to keep track of all the information
  about the initial state. However, if we encode a small message (e.g. with one degree of freedom) in the initial state as in the case of a product state,
  it is likely that the message fits in subsystems with parametrically larger number of degrees of freedom.  
However, a careful look at theorem \ref{theo} shows that this is not necessarily true. 
Intuitively, if our Hamiltonian of interest is very efficient in generating entanglement, there might not be much room left in large subsystems to hold information.
This is reflected in parameter $\delta^\L$ in (\ref{uni}).
Consider the universality argument for initial state with $\psi_{\bar{M}}$ fixed and a typical $\phi_M\in\cH_M$:
  \bea\label{scram}
  &&\la \|\omega_S(M)-\la\omega_S\ra_M\|\ra\leq \sqrt{\frac{d_S^\L\delta^\L}{d_M}},\\
  &&\delta^\L=\sum_k \la k|\frac{P^M}{d_M}|k\ra\: \tr[\left(\tr_{\bS}(P^\L|k\ra\la k|P^\L)\right)^2],
  \eea
 where $|k\ra$ are the eigenstates of the full interacting Hamiltonian, $P^M$ is the projector to $\cH_M$.
 If the right hand side of (\ref{scram}) is small,
   the message $M$ is scrambled in subsystems of size less than or equal to $S$. 
   The parameter $\delta^\L$ is a weighted average of regularized subsystems' purities in energy eigenstates.
   Note that if the subsystem purity (equivalently its second Renyi entropy) is bounded above (below) by an energy independent quantity $\alpha/d_S^\L$, then 
   $\delta^\L \:d_S^\L\leq \alpha$. The logarithm on the right hand side of (\ref{scram}) provides 
   a sufficient condition for scrambling:
   \bea
   n_M-\log\alpha\gg 1,
   \eea
     where $n_A=\log(d_A)$ denotes the number of degrees of freedom in subspace $\cH_A$.
     
   The Renyi entropy of subsystems in vacuum is a topic that has attracted a lot of
   attention recently. The ground state of systems with local interactions seem to have an entanglement entropy proportional to the 
   area of the subsystem; see \cite{eisert2010colloquium} for a review. This goes with the name area law, and supports the simple picture that due to the locality of 
   interactions, it is only the degrees of freedom near the boundary that matter in the calculation of entanglement entropy. The same is believed to be
    true for low-energy eigenstates and subsystems' second Renyi entropy. 
   The sufficient condition for scrambling a message $M$  
    in local systems with area law for all energy eigenstates in $d>1$ dimension is:
    \bea
    n_M-n_S-2c\: n_S^{(d-1)/d}\gg 1,
    \eea
    and $c$ is the constant appearing in the area law formula. There is some evidence that in discrete systems there is a high energy sector of the theory that has an extensive 
   entanglement entropy \cite{alba2009entanglement,requardt2006entanglement}. In such cases, the coefficient $c$ plays a crucial role in the size of the largest
    message that is scrambled. It is not clear whether such sectors survive in the continuum limit.
    
   With regards to the scrambling conjecture, the Hamiltonian of the matrix model or other quantum models of black holes 
   include highly non-local interactions. For systems with non-local interactions it is not hard to imagine 
    sectors of the spectrum that have extensive entanglement entropy. It seems to us that finding bounds on the 
    scaling of entanglement entropy of energy eigenstates in  matrix model or other models of quantum black holes is a promising approach to 
    proving that they are scramblers.




\section{Conclusions and discussion}

In this work, we have generalized the recent information-theoretic approach to equilibration to infinite-dimensional Hilbet spaces, including as examples field theories and 
matrix models. We generically find that small subsystems equilibrate, in the sense that their density matrices become 
almost independent of time and initial state. Our results suggest that at higher energies, larger subsystems 
can reach equilibrium. In the limit of infinite energy, we show that any subsystem smaller than half equilibrates.
Our work emphasizes the importance of energy scales for equilibration and scrambling of large subsystems, both in field theory and matrix models. 
It is an interesting fact that even in Hermitian one-matrix model that has $N$ degrees of freedom, the energy-scale relevant for equilibration of 
large subsystems is order $N^2$. This is in contrast with lattice field theories, where the relevant scale is the same order 
as the number of degrees of freedom $N$.

For concrete estimation of the equilibration parameter we focused our attention on a quench problem 
from the high-energy sectors of free theories. However, the formalism is powerful enough to apply to more general ensembles of initial states. The challenge in each case is to find upper bounds on 
the dimensions of the Hilbert space available to subsystems. In our treatment of the problem, we 
 find bounds on the support of the wave-function in the Hilbert space of free theory, and take advantage of our knowledge of the partition function at zero coupling. 
 This is the reason why the dynamical constraint that appears is on the expectation value of 
 the free Hamiltonian. Loosely speaking, the constraint we use, assumes an upper 
 bound on the time average of particle number densities. We believe that this quantity is not a natural observable to consider
 in strongly coupled systems. It would be interesting to reformulate our approach in terms of 
 energy expecation values in the interacting theory, and without any reference to the zero coupling Hilbert space, at least for some simple systems.

 In spite of our powerful results in lattice field theories, we delibrately avoided addressing the equilibration of density matrices corresponding to spatial regions in continuum field
 theories. The missing step in applying typicality to these cases is a faithful\footnote{A truncation that introduces negligible errors.} truncation of the reduced density matrix to a finite-dimensional 
 Hilbert space. Another subtlety appearing in the continuum limit is that one has to introduce a time scale the corresponding to turning on 
 the interaction.\footnote{Turning on the interaction instanteously might introduce infinities in $\bar{m}$. We thank Guy D. Moore for clarifying this issue.} However, this time-scale is clearly negligible compared to the time-scales required for 
 the typicality argument to work.
 
 An interesting question to ask is whether one can generalize our discussion of equilibration to multi-matrix models. 
 In multi-matrix models, the notion of a subsystem, and how the true degrees of freedom interact is obscured by gauge transformation. 
 Our philosophy is that for systems with gauge symmetry, the equilibration is more naturally formulated in terms of
  macro-observables discussed in appendix \ref{coarse} \cite{lashkaritoappear}. 
 
 Our focus in this work was on the equilibration of small and large subsystems by proving upper bounds on the time-averaged trace distance of 
 states from their equilibrium value.  However, in the context of information loss in 
 black holes, one is also interested in deviations from equilibration. Finding a lower bound on the trace distance 
 would be a natural way to address the question of how large is the smallest subsystem that does provably contain 
 information about the full quantum state, but we leave this for future work.
 
\section*{Acknowledgements}
We are grateful to Jens Eisert, Patrick Hayden, Alexander Maloney, Guy D. Moore, Robert Raussendorf, Joan Simon and Mark Van Raamsdonk for valuable conversations.
This work is supported by the National Science and Engineering Research Council of Canada.
\appendix

\section*{Appendices}
\section{Levy's lemma}\label{levy}
\begin{lemma}
 Given a function $f:S^d\to\mathbb{R}$, with Lipschitz constant $\lambda=\sup|\nabla f|$, at any random point $x\in S^d$,
 \bea
 Prob\left[f(x)-\la f\ra_{x\in S^d} \geq \ep\right]\leq 2 e^{-c(d+1)\ep^2/\lambda^2},
 \eea
 where $c=\frac{1}{18\pi^3}$.
\end{lemma}
See \cite{milman1986asymptotic} for a proof.

A pure $d$-dimensional quantum state $|\psi\ra=\sum_k c_k|k\ra$ can be thought of as a point on $S^{2d-1}$ with the real 
and imaginary parts of $c_k$ as coordinates. We apply Levy's lemma to functions $f(\psi)$ to discuss the probability of 
finding $f(\psi)$ far from its ensemble average $\la f(\psi)\ra_{\psi}$.

\section{Coarse-grained observables}\label{coarse}
In a different approach to equilibration, Reimann \cite{reimann2010canonical} showed that, quite generically, the state of quantum systems becomes indistinguishable 
 from its equilibrium with respect to coarse-grained observables. An observable is coarse-grained 
 if it has finite precision, or in other words can have only a finite number of measurement outcomes. 
  
  Following \cite{citeulike:9264197}, we define 
  the distinguishablility of two states $\rho$ and $\sigma$ with respect to a set of coarse-grained observables $\mathcal{M}$ to be:
  \bea
  D_\mathcal{M}(\rho,\sigma)=\frac{1}{2}\sum_r|\tr(M_r\rho)-\tr(M_r\sigma)|,
  \eea
  where $M_r$ are all POVMs that describe the set $\mathcal{M}$. The operational interpretation of distinguishablility is 
  similar to the trace-norm: if $D_{\mathcal{M}}(\rho,\sigma)\leq \ep$ the optimal probability of telling $\rho$ and $\sigma$ apart using observables $\mathcal{M}$ is 
  $\frac{1}{2}(1+\ep)$.
  
  The following theorem was shown in \cite{citeulike:9264197}.
  \begin{theorem}
   Consider a closed quantum system evolving with a Hamiltonian which has non-degenerate energy gaps. The time-averaged 
   distinguishablility of the state of the system from its time-average with respect to a set of observables $\mathcal{M}$ satisfies:
   \bea
   \la D_{\mathcal{M}}(\rho(t),\omega)\ra_t\leq \frac{N(\mathcal{M})}{4\sqrt{d_{eff}(\omega)}}.
   \eea
  \end{theorem}
  where $N(\mathcal{M})$ is the total number of outcomes in $\mathcal{M}$.
  The denominator $d_{eff}(\omega)$ is exponentially large in the number of degrees of freedom. Therefore, for all
   {\it realistic} measurements we expect the state to be indistinguishable from its time-average.

\section{Proof of theorem \ref{theo}:}\label{appC}
The proof we provide here for theorem \ref{theo} closely follows \cite{citeulike:3788627}.
\subsection{Fluctuations:}
\begin{proof}{{\bf of $(i)$:}}
Consider an initial state $\psi_0=\sum_k c_k |k\ra$, where $|k\ra$ is an energy eigenstate of the interacting Hamiltonian after the quench. 
Then, in the absence of degenerate energy gaps \footnote{For a treatment of cases with degenerate gaps see \cite{citeulike:11048022}},
\bea
\rho^\Lambda_S(t)-\omega^\Lambda_S&&=\sum_{k,l} c_kc_l^* \left(e^{-i(E_k-E_l)t}-\la e^{-i(E_k-E_l)t}\ra_t\right)\: \tr_{\bar{S}}\left(P^\L|k\ra\la l|P^\L\right)\nn\\
&&=\sum_{k\neq l} c_kc_l^* e^{-i(E_k-E_l)t}\: \tr_{\bS}\left(P^\Lambda|k\ra\la l|P^\L\right)
\eea
where we have used $\la e^{-i(E_k-E_l)}\ra_t=\delta_{kl}$. From the Cauchy-Schwarz inequality we know that the 
trace norm is bounded above by:
\bea\label{fluc}
\la\|\rho_S^\Lambda(t)-\omega_S^\L\|\ra_t\leq \sqrt{d_S^\L\:\la\tr[(\rho_S^\Lambda-\omega_S^\Lambda)^2]\ra_t}
\eea
The right-hand side of \ref{fluc} can be simplified further:
\bea
\la\tr[(\rho_S^\L-\omega_S^\L)^2]\ra_t&&=\sum_{k\neq l,m\neq n}c_kc_mc_l^*c_n^*\la e^{-i(E_k-E_l+E_m-E_n)t}\ra_t\:\tr\left[\tr_{\bS}\left(P^\L|k\ra\la l|P^\L\right)\tr_{\bS}\left(P^\L|m\ra\la n|P^\L\right)\right]\nn\\
&&=\sum_{k\neq l}|c_k|^2|c_l|^2\tr\left[\tr_{\bS}\left(P^\L|k\ra\la l|P^\L\right)\tr_{\bS}\left(P^\L|l\ra\la k|P^\L\right)\right],\nn\\
&&=\sum_{k\neq l}|c_k|^2|c_l|^2\sum_{s,s',b,b'}\la s b|P^\L|k\ra\la l|P^\L|s'b\ra\la s'b'|P^\L|l\ra\la k|P^\L|sb'\ra\nn\\
&&=\sum_{k\neq l}|c_k|^2|c_l|^2\sum_{s,s',b,b'}\la s b|P^\L|k\ra\la k|P^\L|sb'\ra\la s'b'|P^\L|l\ra\la l|P^\L|s'b\ra\nn\nn\\
&&=\sum_{k\neq l}\tr\left[\tr_S\left(|c_k|^2P^\L|k\ra\la k|P^\L\right)\tr_S\left(|c_l|^2P^\L|l\ra\la l|P^\L\right)\right]\nn\\
&&=\tr[(\omega_{\bS}^\L)^2]-\sum_k|c_k|^4\tr[(\tr_SP^\L|k\ra\la k|P^\L)^2]\nn\\
&&\leq \tr[(\omega_{\bS}^\L)^2]
\eea
 where $\{|s\ra\}$ and $\{|b\ra\}$ are orthonormal bases that span $\cH_S$ and $\cH_{\bS}$, and we have used the absence of degenerate energy gaps: $\la e^{-i(E_k-E_l+E_m-E_n)t}\ra_t=\delta_{kn}\delta_{lm}$. 
 The weak subadditivity of purity \cite{van2002renyi} implies $\tr[(\omega_{\bS}^\L)^2]\leq d_S^\L\tr[(\omega^\L)^2]$. 
Expanding in the orthonormal basis of free theory $\{|{\bf n}\ra\}$ it is clear that 
 \bea
 \tr((\omega^\L)^2)&&=\sum_{{\bf n,m,k,l}}\omega_{nm}\omega_{kl}\la {\bf l}|P^\L|{\bf n}\ra\la{\bf m}|P^\L|{\bf k}\ra\nn\\
 &&=\sum_{{\bf n,m,k,l}\:\in \cH_{E\leq \L}}|\omega_{nm}|^2\nn\\
  &&\leq\sum_{{\bf n,m,k,l}}|\omega_{nm}|^2= \tr(\omega^2).
 \eea
 Putting this back into (\ref{fluc}) we find:
 \bea
 \la\|\rho_S^\Lambda(t)-\omega_S^\L\|\ra_t\leq \sqrt{\frac{(d_S^\L)^2}{d_{eff}(\omega)}}=\eta_S^\L
 \eea
 \qed
 \end{proof}
%
\subsection{Truncation Error}
\begin{proof}{{\bf of $(ii)$:}}
Expand the state $\psi(t)$ in the orthonormal basis of free theory:
\bea
\psi(t)=\sum_{\bf n}c_{\bf n}(t)c_{\bf m}(t)|{\bf n}\ra\la {\bf m}|
\eea
with $|{\bf n}\ra$ denoting the eigenstate $|n_1,\hdots,n_N\ra$. The expectation value of the free Hamiltonian at time $t$ satisfies:
\bea\label{c1}
\tr(\psi(t)H_{free})&&\geq \sum_{{\bf n}:\sum_i\mu_in_i,=\Lambda}^{\infty}|c_{\bf n}(t)|^2\:(\sum_i\mu_in_i)\geq \Lambda  \sum_{{\bf n}:\sum_i\mu_in_i=\Lambda}^{\infty}|c_{\bf n}(t)|^2\nn\\
&&\geq \Lambda\: \tr(\psi-\psi^\Lambda).
\eea
The matrix $\psi(t)= \begin{pmatrix}
  \psi^\Lambda & A \\
  A^\dagger & \psi^{UV}\\
  \end{pmatrix}$ is positive semi-definite, and as a result $\|A\|^2\leq \|\psi^\Lambda\|\|\psi^{UV}\|\leq \|\psi^{UV}\|$ \cite{ohliger2011continuous}.
  From (\ref{c1}) and the triangle inequality we find
  $$\|\psi(t)-\psi^\Lambda(t)\|\leq 2\|A\|+\|\psi^{UV}\|\leq 3\sqrt{\|\psi^{UV}\|}\leq 3\sqrt{\frac{\tr(\psi(t)H_{free})}{\Lambda}}$$
Hence, the truncation error is bounded above by
\bea
\la\|(\rho_S(t)-\omega_S)-(\rho_S^\Lambda(t)-\omega_S^\Lambda)\|\ra_t&&\leq \la\|(\psi(t)-\psi^\Lambda)\|\ra_t+\|\omega-\omega^\Lambda\|\nn\\
&&\leq 6\sqrt{\frac{\tr(\omega H_{free})}{\Lambda}},
\eea
where we have used the fact that partial trace as a quantum operation brings density operators closer, i.e. $\|\rho_S-\sigma_S\|\leq \|\rho-\sigma\|$.
\end{proof}

\subsection{Typicality}
\begin{proof}{{\bf of $(iii)$:}}
We are interested in finding the probability for $d_{eff}(\omega)$ to be small.
Let us define the following two maps:
\bea
&&\bF[\rho]=\sum_k |k\ra\la k|\rho|k\ra\la k|\nn\\
 &&\tbF[\rho]=\sum_k |\tk\ra\la k|\rho|k\ra\la \tk|,
\eea
where $|k\ra$ are eigenstates of the interacting Hamiltonian, and $|\tk\ra=\frac{P_\Delta|k\ra}{\|P_\Delta|k\ra\|}$ are their projections to the initial ensemble.
The first map $\bF$ dephases states by killing off-diagonal elements in the interacting Hamiltonian basis,
e.g. $\omega=\bF[\psi_0]$. The second map $\tbF[\psi]$ acts on states in $\cH_T$ to dephase and projects them back to the initial ensemble $\cH_R$.

An important observation is that the purity of $\tbF[\psi_0]$ is larger than that of $\omega$, and therefore $d_{eff}(\omega)\geq 1/\tr(\tbF[\psi_0]^2)$.
Then, to prove $(iii)$ it suffices to show that 
\bea
\text{Prob}\left(\tr(\tbF[\psi_0]^2)>\frac{4}{d_\Delta}\right)\leq 2 e^{-c\sqrt{d_\Delta}} 
\eea
by applying Levy's lemma to the function
\bea
f(\psi_0)=\ln\left( \tr(\tbF[\psi_0]^2)\right).
\eea
We need to find upper bounds on the average and the Lipschitz constant of $f$. We employ following lemmas:
\begin{lemma}
 For $\psi\in\cH_R$:
 \bea\label{lemm}
 \la|\psi_0\ra\la \psi_0|\otimes |\psi_0\ra\la\psi_0|\ra_{\psi_0}=\frac{(P^\Delta\otimes P^\Delta)\:(\bI+\mathbb{S})}{d_\Delta(d_\Delta+1)}.
 \eea
 The swap operator $\mathbb{S}$ is defined by $\mathbb{S}|l,k\ra=|k,l\ra$, and we have used 
 the trick $\tr(A B)=\tr(\mathbb{S}\:(A\otimes B))$.
  \end{lemma}
\begin{lemma}\label{f(psi)}
The Lipschitz constant of $f(\psi)\equiv\ln\left(\tr\left((\tbF|\psi\ra\la\psi|)^2\right)\right)$ is bounded above according to:
\bea\label{lips}
\lambda=|\nabla f|\leq 4 d_\Delta^{1/4}.
\eea
\end{lemma}
The proofs appear in Appendix B of \cite{citeulike:3788627}. With above lemmas in mind, it is not hard to come by an 
upper bound on $\la f\ra_{\psi_0}$:
\bea\label{ave}
\la f\ra_{\psi_0}&&\leq \ln\la \tr(\tbF[\psi_0]^2)\ra_{\psi_0}\nn\\
&&\leq \ln\tr\left(\mathbb{S}\:(\tbF\otimes\tbF\la|\psi_0\ra\la\psi_0|\ra_{\psi_0})\right)\nn\\
&&=\ln\left(\sum_{k,l}\tr\left(|\tilde{k}\tilde{l}\ra\la k l|\left(\frac{(P^\Delta\otimes P^\Delta)(\bI+\mathbb{S})}{d_\Delta(d_\Delta+1)}\right)|k l\ra\la \tilde{l}\tilde{k}|\right)\right)\nn\\
&&\leq \ln\left(\sum_{k,l}\la\tilde{l}\tilde{k}|\tilde{k}\tilde{l}\ra\left(\frac{\la k l|(P^\Delta\otimes P^\Delta)(|k l\ra +|l k\ra)}{d_\Delta(d_\Delta+1)}\right)\right)\nn\\
&&\leq\ln\left(\frac{2}{d_\Delta(d_\Delta+1)}\sum_{k l}\la l k|P^\Delta\otimes P^\Delta|k l\ra\right)\nn\\
&&\leq \ln\left(\frac{2}{d_\Delta}\right),
\eea

Plugging (\ref{lips}) and (\ref{ave}) into Levy's lemma for $f[\psi_0]$ gives: 
\bea
\text{Prob}\left(\tr(\tbF[\psi_0]^2)>\frac{4}{d_\Delta}\right)\leq 2 e^{-c\sqrt{d_\Delta}},
\eea
which completes the proof of typicality.
\end{proof}

\subsection{Universality}
\begin{proof}{{\bf of $(iv)$:}}
 We apply Levy's lemma to the function $g[\psi_0]=\|\omega_S^\L-\la\omega_S^\L\ra_{\psi_0}\|$. The 
 Lipschitz constant of $g$ satisfies $|\nabla g|\leq 1$. As for the upper bound on $\la g\ra_{\psi_0}$ we use the Cauchy-Schwarz inequality:
\bea\label{4iv}
\la g\ra_{\psi_0}&&\equiv\la\|\omega_S^\L-\la\omega_S^\L\ra_{\psi_0}\|\ra_{\psi_0}\nn\\
&&\leq \sqrt{d_S^\L\:\la\tr[(\omega_S^\L-\la \omega_S^\L\ra_{\psi_0})^2]\ra_{\psi_0}}
\eea
The right hand side of \ref{4iv} is bounded by:
\bea
\la\tr[(\omega_S^\L-\la \omega_S^\L\ra_{\psi_0})^2]\ra_{\psi_0}&&=\tr\left[\mathbb{S}\left(\la\omega_S^\L\otimes\omega_S^\L\ra_{\psi_0}-\la\omega_S^\L\ra_{\psi_0}\otimes \la\omega_S^\L\ra_{\psi_0}\right)\right]\nn\\
&&=\tr_{SS}\left[\mathbb{S}\left(\tr_{\bS\bS}\left[P^\L(\bF)\otimes P^\L(\bF)\left(\la \psi_0\otimes \psi_0\ra_{\psi_0}-\frac{P^\Delta\otimes P^\Delta}{d_\Delta^2}\right)\right]\right)\right]\nn\\
&&=\tr_{SS}\left[\mathbb{S}\left(\tr_{\bS\bS}\left[P^\L(\bF)\otimes P^\L(\bF)\left(\frac{P^\Delta\otimes P^\Delta(\mathbb{I}+\mathbb{S})}{d_\Delta(d_\Delta+1)}-\frac{P^\Delta\otimes P^\Delta}{d_\Delta^2}\right)\right]\right)\right]\nn\\
&&\leq\tr_{SS}\left[\mathbb{S}\left(\tr_{\bS\bS}\l
\left[P^\L(\bF)\otimes P^\L(\bF)\left(\frac{(P^\Delta\otimes P^\Delta) \mathbb{S})}{d_\Delta^2}\right)\right]\right)\right]\nn\\
&&\leq\sum_{lk}\frac{\la kl|P^\Delta\otimes P^\Delta|lk\ra}{d_\Delta^2}\:\tr_{SS}\left[\mathbb{S}\left(\tr_{\bS\bS}\left[P^\L\otimes P^\L| kl\ra\la kl|P^\L\otimes P^\L\right]\right)\right]\nn\\
&&=\sum_{lk}\frac{\la kl|P^\Delta\otimes P^\Delta|lk\ra}{d_\Delta^2}\:\tr_{S}\left[\tr_{\bS}[P^\L| k\ra\la k|P^\L]\:\tr_{\bS}[P^\L|l\ra\la l|P^\L]\right]\nn\\
&&\leq\sum_{lk}\frac{\la k|P^\Delta|l\ra\la l|P^\Delta|k\ra}{2d_\Delta^2}\:\tr_{S}\left[(\tr_{\bS}[P^\L| k\ra\la k|P^\L])^2+(\tr_{\bS}[P^\L|l\ra\la l|P^\L])^2\right]\nn\\
&&=\frac{1}{d_\Delta^2}\sum_k\la k|P^\Delta|k\ra\:\tr_S[(\tr_{\bS}P^\L|k\ra\la k|P^\L)^2]=\frac{\delta^\L}{d_\Delta}
\eea
where we have defined $P^\L(\bF)[\sigma]=P^\L \bF[\sigma]P^\L$ and used \ref{lemm}. Therefore, $\la f\ra_{\psi_0}\leq \sqrt{d_S^\L\delta^\L/d_\Delta}$. Plugging these into Levy's lemma for
 $g(\psi_0)$ completes the proof.
\end{proof}

\section{Asymptotics of restricted partitions}\label{asymp}
The large $E$ asymptotic formula for the number of partitions of integer $E$ is given by 
\bea
 P(E)\simeq \frac{1}{4\sqrt{3}E}\:e^{\pi\sqrt{2E/3}},
 \eea
 which is the first term that appears in the asymptotic expansion obtained by Hardy and Ramanujan in \cite{hardy1918asymptotic}.
 
 The asymptotics of the restricted partition function $Q(E,s)$ defined by the number of partitions $E$ into exactly $s$ parts is studied \cite{knessl1990partition}, and has the following three regimes:
 \begin{enumerate}
  \item \text{If $s=O(1)$ and $E\gg 1$}: $$Q(E,s)\simeq\frac{E^{s-1}}{s! (s-1)!}$$
  \item \text{If $s/\sqrt{E}=O(1)$ and $E\gg 1$}: $$Q(E,s)\simeq\frac{a^2}{2\pi s^2}\frac{\exp\left(\frac{2s}{a}\int_0^a\frac{dt\: t}{e^t-1}-s \log(1-e^{-a})-\frac{a}{2}\right)}{\left((e^a-1)\int_0^a \frac{dt\: t^2 e^t}{(e^t-1)^2}\right)^{1/2}},$$
  where $a$ is found by solving
$$\frac{E}{s^2}=\frac{1}{a^2}\int_0^a\frac{dt\: t}{e^t-1}.$$
\item \text{If $s/\sqrt{E}\gg1$, and $E\gg s\gg 1$}: 
$$Q(E,s)\simeq \frac{1}{4\sqrt{3}E}\exp\left(\pi\sqrt{\frac{2E}{3}}-\frac{\pi\:s}{\sqrt{6E}}-\frac{\sqrt{6E}}{\pi}\:e^{-\frac{\pi s}{\sqrt{6E}}}\right)$$
 \end{enumerate}
  The number of partitions of $E$ into exactly $s$ parts is equal to the number of
  partitions of $E-s$ into at most $s$ parts: $Q(E,s)=P(E-s,s)$. Since we are only interested in the limit $E\gg s$, the asymptotic behaviour of $Q(E,s)$ and $P(E,s)$ are the same.
   In section \ref{higher} we provide a physicist's derivation of these formulae. 
 
%
In number theory, the denumerant $D(E;a_1,\hdots,a_s)$ denotes the number of ways one can partition a positive integer $E$ into integer parts 
$a_1,\hdots, a_s$: $E=\sum_i{n_i}a_i$. The asymptotic behvaiour of the denumerant for large $E$ and $s=O(1)$ is given by
\bea
D(E;a_1,\hdots,a_s)=\frac{E^{s-1}}{(s-1)!\:a_1\hdots a_s}.
\eea
For a reference on asymptotic properties of denumerant see \cite{agnarsson2002sylvester}.
 
\section{Equilibration time}
In this appendix, we would like to elaborate on time-scales relevant for weak equilibration. One way to approach this problem is
 to bound the trace distance $\|\rho_S(t)-\omega_S\|$ averaged over a finite time interval $[0,T]$. This approach was  
 discussed in \cite{citeulike:11048022}, and a generalization of the fluctuation theorem to finite-time averages was obtained. Applied to our regularized Hilbert spaces we find:
  \bea
 \la\|\rho_S^\L(t)-\omega_S^\L\|\ra_T\leq \eta_S^\L\sqrt{1+\frac{8\log_2 d_\L}{\ep T}}
 \eea
 where $\la f\ra_T=\frac{1}{T}\int_0^Tdt f(t)$, $d_\L$ is the dimension of the total Hilbert space $\psi^\L$ explores, and 
 $\ep$ is defined by the minimum spaces between energy gaps:
 \bea
 \ep=\min_{a,b}\{|G_{(a,b)} -G_{(c,d)}|:\text{$(a,b)\neq (c,d)$}\}.
 \eea
The minimum is over all energy gaps $G_{(a,b)}=E_a-E_b$ in the spectrum of the interacting Hamiltonian. We expect 
$\ep$ to be at least as small as $O(d_\L^{-1})$, which gives an equilibration time-scale exponential in the number of degrees of freedom. This time scale 
is extremely long, and does not seem to provide any insights into the dynamics of equilibration. We could have anticipated this based on
  the fact that it takes a long time for $\psi(t)$ to explore the exponentially large Hilbert space needed for the typicality argument to work.
   A stronger bound would require restriction to a specific set of Hamiltonians or further dynamical assumptions.

\bibliographystyle{JHEP}
\bibliography{micro4}
\end{document}